\documentclass[11pt]{article}
  
\usepackage{amsmath,amssymb}

\usepackage{epsfig}  
\usepackage{graphicx}               
\usepackage{url}
\usepackage{hyperref}
\usepackage{slashed}
\usepackage[makeroom]{cancel}
\usepackage{cite}

\usepackage{pstricks}
\usepackage{color}

\usepackage{tikz}
\usetikzlibrary{positioning}
\usetikzlibrary{decorations.pathmorphing}
\usetikzlibrary{decorations.markings}
\usetikzlibrary{snakes}
 
\newcommand{\nc}{\newcommand}  

\nc{\beq}{\begin{equation}}  
\nc{\eeq}{\end{equation}}  
\nc{\beqa}{\begin{eqnarray}}  
\nc{\eeqa}{\end{eqnarray}}  
\nc{\bea}{\begin{eqnarray}}  
\nc{\eea}{\end{eqnarray}}  
\nc{\ra}{\rightarrow}  
\nc{\lsim}{\begin{array}{c}\,\sim\vspace{-21pt}\\< \end{array}}  
\nc{\gsim}{\begin{array}{c}\sim\vspace{-21pt}\\> \end{array}}  
\nc{\Tr}{{\rm Tr}}
\nc{\slsh}{\slash\hspace*{-0.22cm}}

\def\be{\begin{equation}}
\def\ee{\end{equation}}
\def\bea{\begin{eqnarray}}
\def\eea{\end{eqnarray}}
\def\bit{\begin{itemize}}
\def\eit{\end{itemize}}

\def\to{\rightarrow}

\title{  
\vspace*{-2.3cm}  
\begin{flushright}  
\normalsize{  
SLAC-PUB-15445 
  }  
\end{flushright}  
\vspace{1.5cm}  
\Large  
\textbf{
Up Sector of Minimal Flavor Violation: Top Quark Properties and Direct $D$ meson $CP$ violation
}\vspace*{1.0cm}   
}

\author{Yang Bai$^{a}$, Joshua Berger$^{b}$, JoAnne L. Hewett$^{b}$, Ye Li$^{b}$
\vspace{5mm}
\\
$^{a}$ \normalsize\emph{Department of Physics, University of Wisconsin, Madison, WI 53706, USA}  \vspace{1mm} \\
$^{b}$ \normalsize\emph{SLAC National Accelerator Laboratory, 2575 Sand Hill Road, Menlo Park, CA 94025, USA}
}

\date{}

\begin{document}  

\tikzset{ 
  scalar/.style={dashed},
  scalar-ch/.style={dashed,postaction={decorate},decoration={markings,mark=at
      position .5 with {\arrow{>}}}},
  fermion/.style={postaction={decorate}, decoration={markings,mark=at
      position .5 with {\arrow{>}}}},
  gauge/.style={decorate, decoration={snake,segment length=0.2cm}},
  gauge-na/.style={decorate, decoration={coil,amplitude=4pt, segment
      length=5pt}}
}

\setcounter{page}{0}  
\maketitle  

\vspace*{1cm}  
\begin{abstract} 
Minimal Flavor Violation in the up-type quark sector leads to
particularly interesting phenomenology due to the 
interplay of flavor physics in the charm sector and
collider physics from flavor changing processes in the top sector.  We study the most
general operators that can affect top quark properties and $D$ meson
decays in this scenario, concentrating on two $CP$ violating operators
for detailed studies. The consequences of these effective operators on
charm and top flavor changing processes are generically small, but can
be enhanced if there exists a light flavor mediator that is a Standard
Model gauge singlet scalar and transforms under the flavor symmetry
group. This flavor mediator can satisfy the current experimental
bounds with a mass as low as tens of GeV and explain observed
$D$-meson direct $CP$ violation. Additionally, the model predicts a
non-trivial branching fraction for a top quark decay that would mimic
a dijet resonance.
\end{abstract}  
  
\thispagestyle{empty}  
\newpage  
  
\setcounter{page}{1}

\baselineskip18pt   

\vspace{-3cm}

\section{Introduction}
\label{sec:intro}
Low scale extensions of the Standard Model (SM) are forced to contend
with the so-called New Physics Flavor Puzzle (NPFP): new physics
at or below the TeV scale must have non-generic flavor structure to
satisfy experimental constraints.  This problem is
exacerbated by the recent discovery of a SM-like Higgs
boson~\cite{Aad:2012tfa,Chatrchyan:2012ufa}, which lowers the scale of
new physics required to have a natural solution to the hierarchy
problem.  In order to solve the NPFP, a model of new physics must
either be unnatural, with a high scale, or have some mechanism that
strongly suppresses low-energy flavor violating interactions, such as
flavor blindness, alignment or Minimal Flavor Violation (MFV)
\cite{Chivukula:1987py,D'Ambrosio:2002ex} .  Within MFV, the flavor
constraints on the scale of new physics can be reduced from ${\cal
  O}(10^3~\rm{TeV})$ to ${\cal O}(\rm{few}~\rm{TeV})$. While
there are many studies of the down-type quark sector in the MFV
framework
\cite{D'Ambrosio:2002ex,Buras:2001af,Buras:2002wq,Buras:2003td,Bobeth:2005ck,Blanke:2006ig,Blanke:2006yh,Altmannshofer:2007cs,Hurth:2008jc,Buras:2010mh,Batell:2010qw,Hurth:2012jn,Buras:2013raa},
less attention has been paid to the 
up-type sector.  In this paper, we explore several interesting
phenomena of MFV models in the up-type quark sector.  

By treating the SM Yukawa couplings as spurions, MFV provides a
systematic way to classify the effective higher-dimensional
flavor-violating operators.  One can then determine the most important
operators for a given process based on the dimension of the operators
and the Yukawa matrix insertions.  One interesting feature of MFV
is that the same operator can relate flavor-changing process
predictions for one generation to those for another generation, as
has already been observed in the down-type quark sector with the correlation
of $K^0-\overline{K}^0$ and $B^0-\overline{B}^0$ mixings.  Applying 
MFV to the up-type quark sector, the correlations become more interesting
because of the large mass gap between the charm quark and top
quark masses.  Low-energy tests of charm quark flavor violation can be
directly related to top quark properties probed in high-energy
experiments, including the Large Hadron Collider (LHC). The main focus
of this paper is to explore this correlation on phenomena at different
energy scales.

Both $\Delta F=2$ and $\Delta F=1$ processes are predicted from the
MFV operator analysis. In the up-quark sector, generic $\Delta F=2$
operators are severely constrained by $D^0-\overline{D}^0$
mixing~\cite{Golowich:2007ka}. In MFV, the relevant operators for
$D^0$ -- $\overline{D}^0$ mixing are suppressed by both the bottom quark Yukawa
coupling and Cabibbo-Kobayashi-Maskawa (CKM) mixing angles; they are
much more weakly constrained.  Modifications of top quark properties
are therefore suppressed by the cutoff of the effective operator as
well as by CKM mixing angles, although an interesting signature of same-sign
top pairs could be generated at colliders.  Bounds from $D^0$ --
$\overline{D}^0$ mixing are sufficient to constrain $\Delta T = 2$
operators such that no accessible collider phenomenology is allowed.  

This work therefore focuses on $\Delta F = 1$ MFV operators in the
up-type quark sector.  One immediate consequence of such operators is
that the decays of both $D$ mesons and top quarks can be modified.  In
the SM, the decays of these particles are unsuppressed by CKM angles.  If
the new physics operators are generated by integrating out a heavy
particle above a few hundred GeV, the new contributions to these decays
are negligibly small, even if one considers sensitive $CP$ violating
observables.  The story is different if there is a new electroweak and
color singlet particle $\phi$ lighter than the top quark.  Effective operators can
still describe the new physics contribution to $D$-meson
decays, but the cutoff scales of the relevant operators can now
be as low as ${\cal O}(10~\rm{GeV})$.  The top quark, on the other
hand, can directly decay to $\phi$ and a light quark (a
similar decay into a charged Higgs plus $b$-quark has been studied
in Ref.~\cite{Barger:1989fj}).   This new and potentially large
decay channel for the top quark is currently allowed and requires a
dedicated search at the LHC.  The branching ratio and search
strategy could be dramatically different from the effective operator
analysis in Ref.~\cite{Faller:2013gca}.

Light neutral scalars commonly arise as pseudo-Nambu-Goldstone bosons
from spontaneous breaking of a global symmetry and as scalars, which
may be elementary or composite, in some hidden sector (see
Ref.~\cite{Bai:2010kf} for the effects on $B$ physics in this
scenario).  If the global $SU(3)^5$ flavor symmetry in the SM is
spontaneously broken and a small explicit breaking is added,
``light familons"~\cite{Feng:1997tn} are generic predictions,
particularly within the MFV framework.  The existence of a light
$\phi$ charged under the flavor symmetry is well motivated from this
perspective.  In this paper, rather than explore the symmetry
breaking mechanism of the global flavor symmetry, we study the
phenomenological consequences of the light $\phi$ field for $\Delta C
=1$ and $\Delta T =1$ processes.  

This paper is organized as follows. In Section~\ref{sec:operators}, we
first classify all four-fermion $\Delta F = 1$ operators in MFV
involving up-type quarks and introduce the light $\phi$ field that can
generate large coefficients for these operators.  We then study
modifications of top quark properties in
Section~\ref{sec:topquarkproperty}, including single top production,
$t \bar t$ pair-production, and non-standard decays of the top quark.  In
Section~\ref{sec:charmphysics}, we calculate predictions for several
$b$-quark and $c$-quark related observables.  Particular attention
is devoted to direct $CP$ violation of the neutral $D$ meson. The
discussion of UV completions of this model and the conclusions of this work
are presented in Section~\ref{sec:conclusion}.  A calculation of the partial width
$\Gamma(Z \rightarrow q\bar{q}^\prime \phi)$ is discussed in
Appendix~\ref{sec:appa}, running of the relevant
Wilson coefficients in Appendix~\ref{sec:appb}, and hadronic
matrix element estimation in Appendix~\ref{sec:appc}.

\section{Up Sector Operators and Models}
\label{sec:operators}
In practice, the principle of MFV is implemented by treating the SM
Yukawa matrices as spurions of flavor symmetry.  A MFV operator can
then be written down by demanding that it is formally flavor
invariant.  The quark sector before introducing the Yukawa couplings has a global
flavor symmetry
\beqa
G^{\rm EW}_F = SU(3)_{Q_L} \times SU(3)_{u_R} \times SU(3)_{d_R} \times U(1)_B \times U(1)_Y \times U(1)_{PQ} \,,
\eeqa
where $U(1)_B$ is global baryon symmetry, $U(1)_Y$ is gauge
hypercharge symmetry, and $U(1)_{PQ}$ is Peccei-Quinn symmetry. In the
SM, the $U(1)_{PQ}$ is explicitly broken by the Yukawa couplings,
while in MSSM-like two Higgs doublet models, the Yukawa couplings
preserve the $U(1)_{PQ}$ symmetry by assigning opposite charges for
$H_d$ and down-type quarks.  Concentrating on the non-Abelian global
symmetries, the SM Yukawa matrices can be treated as spurions with
representations
\beqa
 Y_U \sim (3, \bar{3}, 1 ) \,, \qquad  Y_D \sim (3, 1, \bar{3} ) \,.
\eeqa
where ``$U$" represents $(u, c, t)$ quarks and ``$D$" represents $(d, s, b)$ quarks.  

The description above is the standard description of MFV.  There is,
however, an equivalent formulation that will be more convenient for
studying particles with mass $m \ll v$, where $v$ denotes the
Higgs Vacuum Expectation Value (VEV).  In this second approach, which we refer to as the
$\cancel{\rm EW}$ approach, we construct operators invariant
only under the $U(1)_{EM}$ subgroup of the electroweak gauge group
$SU(2)_L \times U(1)_Y$.  Any UV completion will, of course, generate
$SU(2)_L \times U(1)_Y$ invariant operators, but in the limit we are
considering, we can postpone such high energy considerations.  One
consequence of the $\cancel{\rm EW}$ approach is that the left-handed
quark fields can be rotated separately.  Couplings to the $W$ boson
provide additional flavor violation.  The flavor structure can be
described by the group
\beqa
G^{\cancel{\rm EW}}_F = SU(3)_{u_L} \times SU(3)_{d_L} \times SU(3)_{u_R} \times SU(3)_{d_R} ,
\label{eq:flavor-symmetry-EW-broken}
\eeqa
under which there are spurions
\beqa
 \lambda_U \sim (3, 1,  \bar{3}, 1 ) \,, \quad \lambda_D \sim (1, 3,  1, \bar{3} )  \,,  \quad
 V  \sim (3, \bar{3}, 1, 1)  \,.
\eeqa
We are free to choose a basis where $\lambda_U = \rm{diag}\{
\lambda_u, \lambda_c, \lambda_t \}$, $\lambda_D =\rm{diag} \{
\lambda_d, \lambda_s, \lambda_b \}$, and $V$ is the CKM
matrix.  Up to ${\cal O}( \lambda_U^2, \lambda_D^2, \lambda_D
\lambda_U)$, we present all MFV $\Delta F = 1$ operators in
Table~\ref{tab:effective-operators}.\footnote{We neglect operators constructed of $\sigma^{\mu\nu}$,
which either have zero hadronic matrix elements for the $D$ meson decays in the naive
factorization approximation or can be related to scalar operators
via a Fierz transformation.}  None of these operators generate
$\Delta F = 2$ observables at leading order, but, in the models
considered below, such observables will be generated at one-loop.
We will find that constraints from $\Delta F = 2$ processes are nevertheless weak.  For completeness, we also include the
four-fermion operators containing leptons and at least one up-type
quark.  In this case, the MFV structure is analogous to that discussed
above with the replacement $V \to U$, where $U$ is
the Pontecorvo-Maki-Nakagawa-Sakata
(PMNS)~\cite{Pontecorvo:1957cp,Maki:1962mu} matrix.



\begin{table}[!tb]
 \centering
  \renewcommand{\arraystretch}{1.3}
  \begin{tabular}{l}
    \hline \hline
    Operator  \\
    \hline \hline
    $(\overline{u}_{L\,\alpha}\,V\, \gamma^\mu\, d_{L\,\alpha})
    (\overline{d}_{L\,\beta}\, V^\dagger\, \gamma_\mu\, u_{L\,\beta})$ 
    \\
    $(\overline{u}_{L\,\alpha}\, V\, \gamma^\mu \, d_{L\,\beta})
    (\overline{d}_{L\,\beta}\, V^\dagger\, \gamma_\mu\, u_{L\,\alpha})$ 
    \\
    $(\overline{u}_{R\,\alpha} \,\lambda_U^\dagger\, V\,d_{L\,\alpha})
    (\overline{d}_{L\,\beta}\, V^\dagger\, \lambda_U \,u_{R\,\beta})$ \\
    $(\overline{u}_{R\,\alpha}\, \lambda_U^\dagger \,V d_{L\,\beta})
    (\overline{d}_{L\,\beta} \,V^\dagger \,\lambda_U \,u_{R\,\alpha})$ \\
    $(\overline{u}_{L\,\alpha} \,V \,\lambda_D\, d_{R\,\alpha})
    (\overline{d}_{R\,\beta} \,\lambda_D^\dagger\, V^\dagger \, u_{L\,\beta})$  \\
    $(\overline{u}_{L\,\alpha} \,V \,\lambda_D\, d_{R\,\beta})
    (\overline{d}_{R\,\beta} \,\lambda_D^\dagger \,V^\dagger
    \,u_{L\,\alpha})$  \\
    $(\overline{u}_{L\,\alpha} \, V \,\lambda_D\,\lambda_D^\dagger\,
    V^\dagger\,  \gamma^\mu \, u_{L\,\alpha})
    (\overline{u}_{R\,\beta} \, \gamma_\mu\, u_{R\,\beta})$  \\
    $(\overline{u}_{L\,\alpha}\,V \,\lambda_D\,\lambda_D^\dagger\,
    V^\dagger\,  \gamma^\mu \, u_{L\,\beta})
    (\overline{u}_{R\,\beta} \,\gamma_\mu
    \,u_{R\,\alpha})$  \\
    $(\overline{u}_{L\,\alpha} \,V \,\lambda_D\,\lambda_D^\dagger\,
    V^\dagger\,\gamma^\mu \,  u_{L\,\alpha})
    (\overline{d}_{R\,\beta} \, \gamma_\mu\, d_{R\,\beta})$  \\
    $(\overline{u}_{L\,\alpha}\,V \,\lambda_D\,\lambda_D^\dagger\,
    V^\dagger\, \gamma^\mu \, u_{L\,\beta})
    (\overline{d}_{R\,\beta} \,\gamma_\mu
    \,d_{R\,\alpha})$  \\   
    $(\overline{u}_{L\,\alpha}\,V\, \gamma^\mu\, d_{L\,\alpha})
    (\overline{e}_{L}\, U^\dagger\, \gamma_\mu\, \nu_{L})$ \\
            $(\overline{u}_{L\,\alpha}\,V\,\lambda_D\,\lambda_D^\dagger\,V^\dagger\, \gamma^\mu\, u_{L\,\alpha})
    (\overline{e}_{L}\, \gamma_\mu\, e_{L})$  \\
                $(\overline{u}_{L\,\alpha}\,V\,\lambda_D\,\lambda_D^\dagger\,V^\dagger\, \gamma^\mu\, u_{L\,\alpha})
    (\overline{\nu}_{L}\, \gamma_\mu\, \nu_{L})$  \\
        $(\overline{u}_{L\,\alpha}\,V\,\lambda_D\,\lambda_D^\dagger\,V^\dagger\, \gamma^\mu\, u_{L\,\alpha})
    (\overline{e}_{R}\, \gamma_\mu\, e_{R})$  \\
          \hline
      \hline
  \end{tabular}
  \hspace{1.0cm}
  \begin{tabular}{ll}
    \hline \hline
    Operator & Name \\
    \hline \hline
    $(\overline{u}_{L\,\alpha}\,\gamma^\mu\, V\, d_{L\,\alpha})
    (\overline{d}_{R\,\beta} \,\gamma_\mu\, \lambda_D^\dagger\, V^\dagger\, \lambda_U \, u_{R\,\beta})$ & ${\cal O}_{V1}$ \\
    $(\overline{u}_{L\,\alpha}\,\gamma^\mu\, V \,d_{L\,\beta})
    (\overline{d}_{R\,\beta}\, \gamma_\mu\, \lambda_D^\dagger \,V^\dagger\, \lambda_U \,u_{R\,\alpha})$ & ${\cal O}_{V2}$ \\
    $(\overline{u}_{R\,\alpha} \,\lambda_U^\dagger\, V\, d_{L\,\alpha})
    (\overline{d}_{R\,\beta}\, \lambda_D^\dagger\, V^\dagger \,u_{L\,\beta})$ & ${\cal O}_{S1}$ \\
    $(\overline{u}_{R\,\alpha} \,\lambda_U^\dagger\, V \,d_{L\,\beta})
    (\overline{d}_{R\,\beta} \,\lambda_D^\dagger\, V^\dagger \,u_{L\,\alpha})$ &
    ${\cal O}_{S2}$ \\
     $(\overline{u}_{L\,\alpha}\,V\, \lambda_D\, d_{R\,\alpha})
    (\overline{e}_{R}\, \lambda_E^\dagger U^\dagger\,\nu_{L})$ & $~$ \\
     $(\overline{u}_{R\,\alpha}\,\lambda_U^\dagger V\, d_{L\,\alpha})
    (\overline{e}_{R}\, \lambda_E^\dagger U^\dagger \nu_{L})$ & $~$ \\
         $(\overline{u}_{L\,\alpha}\,V\, \lambda_D\, d_{R\,\alpha})
    (\overline{e}_{L}\, U^\dagger\lambda_\nu^\dagger \nu_{R})$ & $~$ \\
     $(\overline{u}_{R\,\alpha}\,\lambda_U^\dagger V\, d_{L\,\alpha})
    (\overline{e}_{L}\,U^\dagger \lambda_\nu^\dagger \nu_{R})$ & $~$ \\
    \hline
      \hline
  \end{tabular}
  \caption{A complete list of four-fermion operators mediating $\Delta F = 1$ processes at the order of ${\cal O} (\lambda^2)$ and satisfying the global symmetry in Eq.~(\ref{eq:flavor-symmetry-EW-broken}). Here, $\alpha$ and $\beta$ are QCD indices. The flavor indices are contracted inside the parenthesis. The operators above in the left panel are Hermitian operators, while the operators in the right panel including ${\cal O}_{V1}$, ${\cal O}_{V2}$, ${\cal O}_{S1}$, ${\cal O}_{S2}$ are all complex and can have $CP$ violating coefficients.
  }
  \label{tab:effective-operators}
\end{table}

The operators in the left panel of Table~\ref{tab:effective-operators}
are Hermitian.  They cannot yield new $CP$ violating phases and can
only violate $CP$ via the CKM phase.  The operators on the right panel,
on the other hand, may have coefficients containing new $CP$ violating
phases. This is unsurprising,  since the matrix $\lambda_D$ may have a
different overall phase from the matrix $\lambda_U$. The operators
${\cal O}_{V1}$ and ${\cal O}_{V2}$ can be rewritten with a scalar
Lorentz structure using a Fiertz transformation as
\beqa
{\cal O}_{V1} &=& 2\,V_{il} \, (\lambda_D^\dagger V^\dagger \lambda_U)_{kj}\,  (\bar{u}^i_{L\,\alpha} u^j_{R\,\beta} ) (\bar{d}^k_{R\,\beta}  d^l_{L\,\alpha}) 
\,, \\
{\cal O}_{V2} &=& 2\,V_{il} \, (\lambda_D^\dagger V^\dagger \lambda_U)_{kj}\,  (\bar{u}^i_{L\,\alpha} u^j_{R\,\alpha} ) (\bar{d}^k_{R\,\beta}  d^l_{L\,\beta}) 
\,,
\eeqa
Similarly, the operators ${\cal O}_{S1}$ and ${\cal O}_{S2}$ can be rewritten as
\beqa
{\cal O}_{S1} &=& \frac{1}{2}\,(\lambda_U^\dagger V)_{il}\,(\lambda_D^\dagger V^\dagger)_{kj}\,  (\bar{u}^i_{R\,\alpha} u^j_{L\,\beta} ) (\bar{d}^k_{R\,\beta}  d^l_{L\,\alpha}) \,+\, \cdots \,, \\
{\cal O}_{S2} &=& \frac{1}{2}\,(\lambda_U^\dagger V)_{il}\,(\lambda_D^\dagger V^\dagger)_{kj}\,  (\bar{u}^i_{R\,\alpha} u^j_{L\,\alpha} ) (\bar{d}^k_{R\,\beta}  d^l_{L\,\beta}) \,+\, \cdots \,,
\eeqa
where, we have not written the tensor operators containing
$\sigma^{\mu\nu}$.  We introduce neutral scalars to UV complete the
operators ${\cal O}_{V2}$ and ${\cal O}_{S2}$ toward the end of this
section.

The remainder of this paper concentrates on the two $CP$-violating operators, ${\cal O}_{V2}$ and
${\cal O}_{S2}$, since these operators are the only ones with a different color
structure compared to the SM that can
contain a new $CP$ violating phase under the assumption of MFV.  They
correspond to the electroweak-invariant operators:
\beqa
{\cal O}^{\rm{EW}}_{V2} &=& 2\, V_{il} \, (\lambda_D^\dagger V^\dagger \lambda_U)_{kj}\,  (\tilde{H}\,\bar{Q}^i_{L\,\alpha} u^j_{R\,\alpha} ) (\bar{d}^k_{R\,\beta}  Q^l_{L\,\beta} H^\dagger)  \,,\\
{\cal O}^{\rm{EW}}_{S2} &=& \frac{1}{2}\,(\lambda_U^\dagger V)_{il}\,(\lambda_D^\dagger V^\dagger)_{kj}\,  (\bar{u}^i_{R\,\alpha} Q^j_{L\,\alpha} ) (\bar{d}^k_{R\,\beta}  Q^l_{L\,\beta})  \,,
\eeqa
where the $SU(2)_L$ indices are contracted in the parenthesis for the
dimension 8 operator ${\cal O}^{\rm{EW}}_{V2}$ and between the two
$Q_L$'s for the dimension 6 operator ${\cal O}^{\rm{EW}}_{S2}$.

\subsection{Phenomenology of the effective operators}
\label{sec:phenoEFT}
We now outline the most relevant $\Delta C =1$ and $\Delta T =1$
processes and perform some preliminary calculations of new physics
contributions using the effective operators ${\cal O}_{V2}$ and ${\cal
  O}_{S2}$.  These contributions are, as will be explored in Sections \ref{sec:topquarkproperty} and
\ref{sec:charmphysics}, the most significant ones.  Keeping only the leading terms in the Lagrangian, we have
\beqa
{\cal O}_{V2}: && \Delta C = 1: \quad 2 \lambda_s \lambda_c V_{12} V^*_{22} (\bar{u}_{L\,\alpha} c_{R\,\alpha} ) ( \bar{s}_{R\,\beta} s_{L\,\beta}) \,,\label{eq:v2domc}  \\
&& \Delta T = 1: \quad 2 \lambda_b \lambda_t V_{11} V^*_{33} (\bar{u}_{L\,\alpha} t_{R\,\alpha} ) ( \bar{b}_{R\,\beta} d_{L\,\beta}) \,,  \label{eq:v2domt}\\
{\cal O}_{S2}: && \Delta C = 1: \quad \frac{1}{2} \lambda_s \lambda_c V_{22} V^*_{12} (\bar{c}_{L\,\alpha} u_{R\,\alpha} ) ( \bar{s}_{R\,\beta} s_{L\,\beta}) \,, \label{eq:s2domc} \\ 
&& \Delta T = 1: \quad 
\frac{1}{2} \lambda_b \lambda_t V_{32} V^*_{33} (\bar{t}_{L\,\alpha} c_{R\,\alpha} ) ( \bar{b}_{R\,\beta} b_{L\,\beta})
 \,. \label{eq:s2domt}
\eeqa
Other terms such as $2 \lambda_b \lambda_t V_{31} V^*_{13}
(\bar{t}_{L\,\alpha} u_{R\,\alpha} ) ( \bar{b}_{R\,\beta}
d_{L\,\beta})$ for the $\Delta T =1$ processes from ${\cal O}_{S2}$
have additional Cabbibo suppression and do not contribute
significantly to the relevant processes.  From
Eqs.~(\ref{eq:v2domc} -- \ref{eq:s2domt}), one  
can see that for a same size contribution to $\Delta C = 1$ processes,
the operator ${\cal O}_{V2}$ has a much larger contribution to $\Delta
T = 1$ processes than the operator ${\cal O}_{S2}$.

We first consider contributions to $D$ meson direct $CP$ violation
from these $\Delta C = 1$ operators.  At this point, we only perform
estimates in an attempt to determine the scale required for
there to be experimentally accessible contributions.  A more complete
analysis is performed in Sec.\ \ref{sec:charmphysics}.  For the decay channel $D^0
(\overline{D}^0) \rightarrow K^+ K^-$, one can define the following
direct $CP$ violating observable
\beqa
A_{CP}^K = \frac{\Gamma(D^0 \rightarrow K^+ K^-) - \Gamma(\overline{D}^0 \rightarrow K^+ K^-) }{\Gamma(D^0 \rightarrow K^+ K^-) + \Gamma(\overline{D}^0 \rightarrow K^+ K^-)} \,.
\eeqa
Assuming maximal $CP$ violating strong and weak phases, the contribution from ${\cal O}_{V2}$ is estimated to be
\beqa
A_{CP}^K \sim \frac{4\sqrt{2} \lambda_c \lambda_s}{N_c G_F} \,\frac{1}{|\Lambda^2|} \approx 0.002 \times \left(\frac{10~\rm{GeV}}{|\Lambda|}\right)^2 \,,
\eeqa
where $|\Lambda|$ is the magnitude of the cutoff of the operator and
the Yukawa coupling values of $y_s$ and $y_c$ are evaluated at the
scale of $M_Z$~\cite{Xing:2007fb}. A similar estimate can be done
for the asymmetry, $A_{CP}^\pi$, involving the decay $D^0 (\overline{D}^0) \rightarrow
\pi^+ \pi^-$, which is highly suppressed by the $d$-quark Yukawa
coupling.  The cutoff must be ${\cal O}$(10~GeV) to generate $A_{CP}^K$
close to the current experimental sensitivity.  

We now perform some preliminary calculations of the most relevant top
quark observable, the single top quark production cross-section.  The
new physics contributions to this observable due to the quark-level
process $u \bar{d} \rightarrow t \bar{b}$ are 
calculated by assuming a scalar particle with a mass $\Lambda$
propagates in a $t$-channel diagram with couplings to $u\bar{t}$
and $d\bar{b}$ denoted by $\bar{\kappa}_U$ and $\bar{\kappa}_D$. There is no CKM
suppression for these couplings, but they are proportional to
appropriate Yukawa couplings.  The Yukawa couplings are, in turn,
sensitive to $\tan\beta$ in two Higgs doublet models.  With some  $\tan \beta$ enhancement,
both couplings $\bar{\kappa}_U$ and $\bar{\kappa}_D$ can be $\mathcal{O}(1)$. Choosing
$\bar{\kappa}_U=\bar{\kappa}_D=1$ and the scalar particle mass to be 10 GeV, we have
new physics contributions to single top production given by
\begin{align}
\sigma[p\bar{p} \rightarrow t \bar{b} (b \bar{t})] &=
0.11~\rm{pb}\qquad (1.96~{\rm TeV}~{\rm Tevatron}), \\
\sigma[p p \rightarrow t \bar{b} (b \bar{t})] &=
0.56~\rm{pb}\qquad (8~{\rm TeV}~{\rm LHC})\,.
\end{align}
Since the contribution to $D$ meson $CP$ violation is linear in the
product of couplings $\bar{\kappa}_U \bar{\kappa}_D$, while top production cross
sections are quadratic in the product of couplings $\bar{\kappa}_U^2
\bar{\kappa}_D^2$, increasing both the mediator mass and couplings
simultaneously keeps contributions to $D$ meson $CP$ violation fixed
while dramatically enhancing single top production.

Similarly, for the other operator ${\cal O}_{S2}$, we have 
\beqa
A_{CP}^K \sim \frac{\sqrt{2}\,\lambda_c \lambda_s}{N_c\,G_F}  \,\frac{\chi_K}{8 N_c}
\frac{1}{\Lambda^2} \approx 0.001 \times \left(\frac{5~\rm{GeV}}{\Lambda}\right)^2  \,,
\eeqa
where the chiral factor $\chi_K \approx 2 m_K^2/(m_c m_s)\approx 4.2$
for $m_c$ and $m_s$ evaluated at the $D$ meson mass.  The single
top production cross sections are suppressed by the CKM element
$|V_{32}|^2$.  For $\bar{\kappa}_U=\bar{\kappa}_D=1$ and scalar
particle mass 10 GeV, they are 
\begin{align}
\sigma[p\bar{p} \rightarrow t \bar{b} (b \bar{t})] &= 5.1\times
10^{-8}~\rm{pb} \qquad (1.96~{\rm TeV}~{\rm Tevatron}) \,, \\
\sigma[p p \rightarrow t \bar{b} (b \bar{t})] &= 3.1\times
10^{-6}~\rm{pb}\qquad (8 {\rm TeV}~{\rm LHC})\,.
\end{align}
This operator has a negligible new physics contribution to the single
top production cross section if there is a sub-percent level
contribution to the $D$ meson $CP$ violation.  From these estimates
for ${\cal O}_{V2}$ and ${\cal O}_{S2}$, one can already see that
different structures of the effective operators that give
contributions to $CP$ violation in $D$-meson decays
have dramatically different predictions for top quark physics.  

\subsection{Light mediators}
\label{sec:mediator}
Because of the Yukawa coupling suppression in MFV, the new physics
affects on $D$ meson $CP$ violation and top quark properties are
typically small for $\Lambda \gtrsim 100$~GeV.  On the other hand, if
the new particle inducing the MFV operator is lighter than $100$~GeV,
then there can be large effects. We do not consider masses below
${\cal O}(10~\rm{GeV})$ because of potentially severe constraints
from decays of bottom-quark bound states and searches for light
hadronic resonances in fixed-target
experiments~\cite{Kephart:1977ad}. For this range of scales, ${\cal
  O}(10~\rm{GeV}) < \Lambda < {\cal O}(100~\rm{GeV})$, $D$ meson
$CP$ violation effects can still be calculated in an effective
operator approach.  For the top quark physics, however, the actual new physics
degree of freedom enters both production and decay.  We therefore pay
special attention to the case in which the MFV operator is generated
by a light particle.  Because of various collider constraints, new
light particles with a mass below 100 GeV must be neutral under SM
gauge interactions.  The particle must be a boson in order to generate
the operators $\mathcal{O}_{V2}$ and $\mathcal{O}_{S2}$.  We introduce
a new scalar gauge singlet $\phi$ which transforms under flavor as a completion for
these operators between the $10$ -- $100~{\rm GeV}$ scales and study
the class of such scalars.

There are several possible flavor representations for the scalar
$\phi$.  We list the flavor symmetry 
possibilities in Table~\ref{tab:scalarreps}.  Note that fields $\phi$ transforming
under $SU(3)_{u_L}$ and under $SU(3)_{d_L}$ are equivalent up to a
basis change since, in either case, the representation under
$G^{\cancel{\rm EW}}_F$ arises from a representation under $G^{\rm EW}_F$.  We therefore
choose, without loss of generality, to consider only cases where
$\phi$ transforms under $SU(3)_{d_L}$.
\begin{table}[htb!]
  \centering
     \renewcommand{\arraystretch}{1.3}
  \begin{tabular}{ccc}
    \hline \hline
    Model \# & $\phi$ flavor & Operators Generated\\
    \hline
    1 & $(1, 3,1,\overline{3})$ & ${\cal O}_{S2}$, ${\cal O}_{V2}$ \\
    2 & $(1, 3,\overline{3},1)$ & ${\cal O}_{S2}$, ${\cal O}_{V2}$ \\
    3 & $(1, 1,3,\overline{3})$ & ${\cal O}_{S2}$ \\
    4 & $(1, 8,1,1)$ & ${\cal O}_{S2}$, ${\cal O}_{V2}$ \\
    5 & $(1, \overline{3},\overline{3},\overline{3})$ & ${\cal O}_{V2}$ \\
    6 & $(1, 8 ,3,\overline{3})$ & ${\cal O}_{S2} $ \\
    7 & $(1, 6, \overline{3},\overline{3})$ & ${\cal O}_{V2}$\\
    \hline \hline
  \end{tabular}
  \caption{Scalar flavor structures under $SU(3)_{u_L}\times
    SU(3)_{d_L}\times SU(3)_{u_R}\times SU(3)_{d_R}$ that are allowed by
    requiring a neutral scalar that reproduces the flavor structure of
    the $CP$-violating operators. 
    } 
  \label{tab:scalarreps}
\end{table}
Some representations can only generate one operator, while other
representations can generate two operators. We will use the
representation $(1, 3, 1, \bar{3})$ as a prototype for our phenomenological
studies, as this representation contributes to both $CP$ violation in
$D$ meson decays and processes involving top quarks, capturing the
full breadth of potential effects due to a light scalar.  At the order of
magnitude level, contributions to a given operator due to the other
representations are comparable.  The leading couplings for a scalar with
this flavor representation are 
\beqa
{\cal L} &\supset& \kappa_{U_L}\, \overline{u}_R^i (\lambda_U^\dagger V)^{il} 
 \phi_{lk} (\lambda_D^\dagger V^\dagger)^{kj} u_L^j
\,+\, \kappa_{U_R}\, \overline{u}_L^i V^{il} \phi_{lk} (\lambda_D^\dagger V^\dagger \lambda_U)^{kj} \,u_R^j 
\,+\, {\rm h.c.} \nonumber \\
&&\,+\, \kappa_D\,\overline{d}_R^k (\phi^\dagger)_{kl} d_L^l \,+\,
{\rm h.c.} \,+\, m_\phi^2 \phi^\dagger \phi \,,
\label{eq:phiInteraction}
\eeqa
where we neglect additional scalar potential terms as well as
additional mass terms that split the $\phi_{lk}$ components.
Perturbativity limits are saturated when $\kappa_D \sim \sqrt{4\pi}$
or $\kappa_{U_L}, \kappa_{U_R} \sim 200$ (the largest coupling is
proportional to $\lambda_b$ and $\lambda_t$).  Under the assumption
that the potential for $\phi$ conserves flavor, the interactions
Eq.~\eqref{eq:phiInteraction} break the global $U(1)$ under which only
$\phi$ transforms.  Redefinitions of $\phi$ then give the freedom to
remove the phase of one of the three couplings.  For concreteness and
without loss of generality, we work in a basis
where $\kappa_D$ is real and the couplings $\kappa_{U_{L, R}}$ are
complex. Integrating out $\phi$, one can generate both
${\cal O}_{V2}$ and ${\cal O}_{S2}$ operators with low energy Lagrangian
\beq
{\cal L} = \frac{\kappa_{U_R} \kappa_D}{2\,m_\phi^2} {\cal O}_{V2}  +
\frac{\kappa_{U_L} \kappa_D}{2\,m_\phi^2} {\cal O}_{S2} + {\rm h.c.}
\eeq
In the following sections, we elaborate on the phenomenology of this model.

\section{Top Quark Properties}
\label{sec:topquarkproperty}
A light $\phi$ with interactions given in
Eq.~(\ref{eq:phiInteraction}) would modify top production and
decay.  From this equation, the leading
couplings mediating $\Delta T =1$ processes are
\beqa
{\cal L} &\supset& \kappa_{U_L} \lambda_b \lambda_t \left( \bar{t}_R \phi_{33} V^*_{33} V^*_{23} c_L
\,+\, \bar{t}_R \phi_{33} V^*_{33} V^*_{13} u_L \right) \,+\,
 \kappa_{U_R} \lambda_b \lambda_t \left( \bar{u}_L \phi_{13} V_{11} V^*_{33} t_R
\,+\, \bar{c}_L \phi_{23} V_{22} V^*_{33} t_R
 \right) \,+\, {\rm h.c.}  \nonumber \\
 &=& \bar{\kappa}_{U_L} \left( \bar{t}_R \phi_{33} V^*_{33} V^*_{23} c_L
\,+\, \bar{t}_R \phi_{33} V^*_{33} V^*_{13} u_L \right) \,+\,
 \bar{\kappa}_{U_R}\left( \bar{u}_L \phi_{13} V_{11} V^*_{33} t_R
\,+\, \bar{c}_L \phi_{23} V_{22} V^*_{33} t_R
 \right) \,+\, {\rm h.c.} \, ,
\eeqa
where, for convenience, we have defined $\bar{\kappa}_{U_L} \equiv
\kappa_{U_L} \lambda_b \lambda_t$ and  $\bar{\kappa}_{U_R} \equiv
\kappa_{U_R} \lambda_b \lambda_t$.  The coupling $\bar{\kappa}_{U_L}$
has an additional CKM angle suppression such that, for perturbative
couplings, there is negligible effect on top physics.  We therefore
focus on contributions from $\bar{\kappa}_{U_R}$. 

\subsection{Single top production}
\begin{figure}[ht!]
\begin{center}
\includegraphics[width=0.45\textwidth]{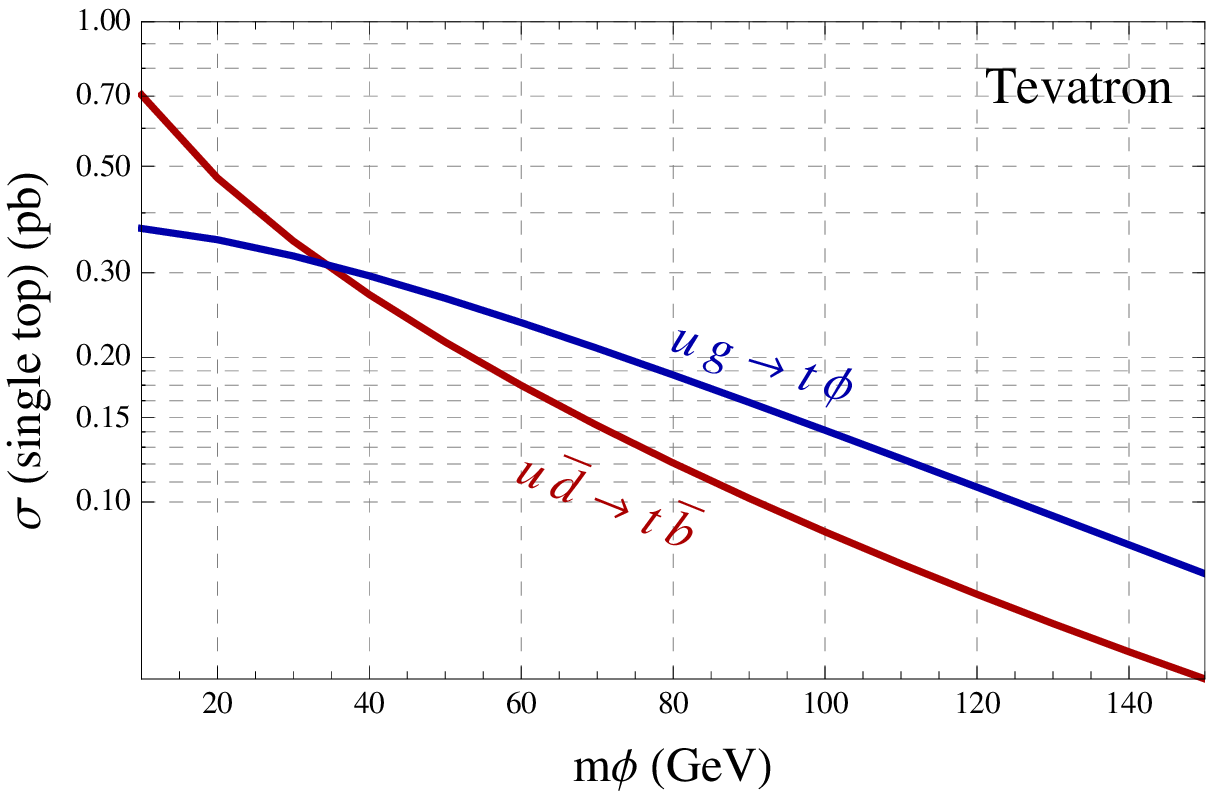} 
\hspace{4mm}
\includegraphics[width=0.46\textwidth]{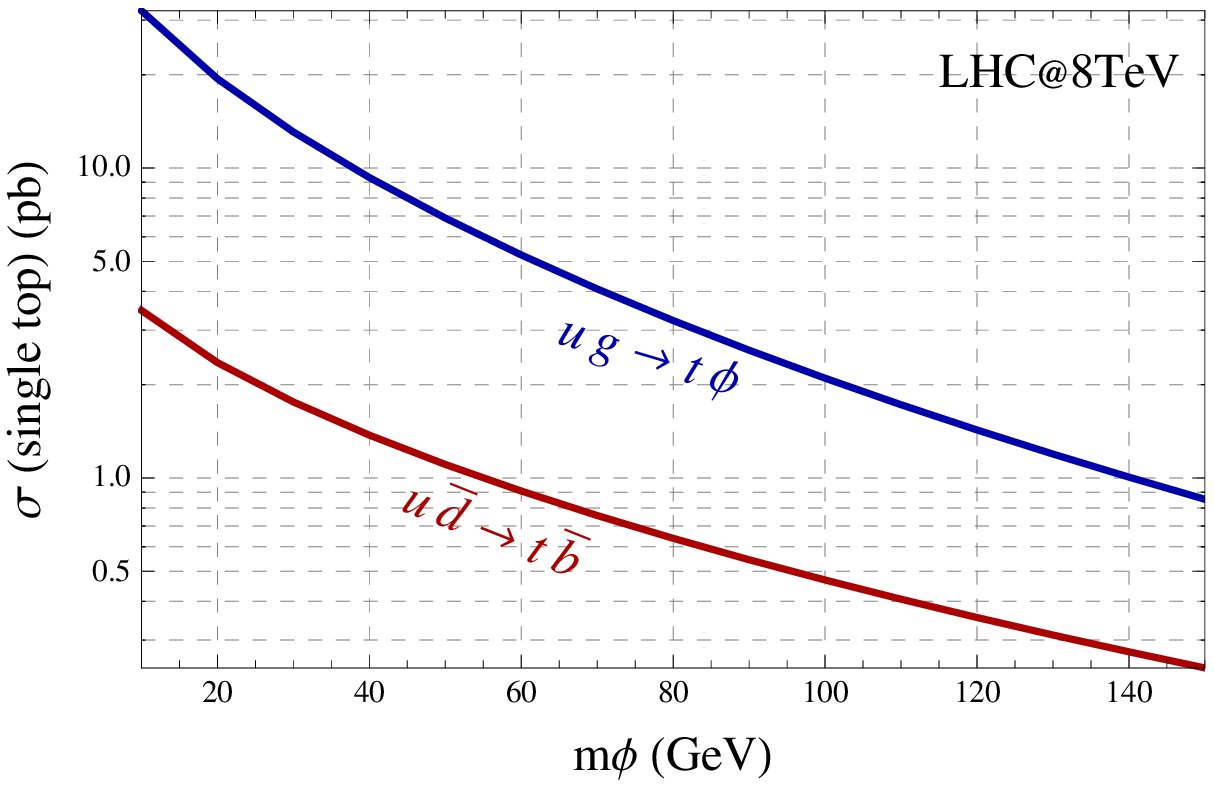} 
\caption{The single top production cross sections from the contributions of the new neutral scalar $\phi$. The couplings are chosen to be $|\bar{\kappa}_{U_R}| = |\kappa_D| = 0.2$. 
}
\label{fig:singletopx}
\end{center}
\end{figure}

In this model, there are two additional contributions to single top production.  The
first is $u\bar{d} \rightarrow t \bar{b}$ with
$\phi_{13}$ being exchanged in the $t$-channel and the second is $u g \rightarrow
t \phi_{13}$ for which there are two tree level diagrams.  For the
first channel, neglecting the $b$-quark mass in the final state, there
is no interference terms between the $W$ mediated and the $\phi_{13}$
mediated diagrams. The leading order contribution to the parton-level
cross section is
\beqa
\sigma (u \bar{d} \rightarrow t \bar{b}) &=& \frac{ |\bar{\kappa}_{U_R}|^2 |\kappa_{D}|^2}{16\pi s^2}  \left[ (2 m_\phi^2  - m_t^2) \ln \left( \frac{m_\phi^2}{s + m_\phi^2 - m_t^2} \right)
+ \frac{ (s - m_t^2) (s + 2 m_\phi^2 - 2 m_t^2 )  }{s + m_\phi^2 - m_t^2} 
\right] \,,
\eeqa
where $\sqrt{s}$ is the parton center-of-mass energy.  There are two diagrams
contributing to the production of $u g \rightarrow t \phi_{13}$: one
is from exchanging a $t$ quark in the $t$-channel and other is from
exchanging a $u$ quark in the $s$-channel. Neglecting the $\phi$
particle mass, the leading order parton-level cross-section is
\beqa
\sigma(u g \rightarrow t \phi_{13}) = \frac{g_s^2\, |\bar{\kappa}_{U_R}|^2 }{192\pi s^3} 
\left[ 2 ( s^2 + 2 s m_t^2 + 2 m_t^4 ) \ln{\left( \frac{s}{m_t^2}  \right)} + 7 m_t^4 - 4 s m_t^2 - 3 s^2 
\right] \,.
\eeqa
The production cross sections at the Tevatron and 8 TeV LHC as a
function of $m_\phi$ are shown in Fig.~\ref{fig:singletopx}.  We use the
Mathematica MSTW 2008 PDFs~\cite{Martin:2009iq}.
Production from $u$ and $g$ partons dominates at the
8 TeV LHC. The latest measurement of single top production from the CDF collaboration at
Tevatron with 7.5 fb$^{-1}$ has an error below 1
pb~\cite{CDF-Single-Top}, while the latest result from the CMS
collaboration with 5.0 fb$^{-1}$ at the 8 TeV LHC has an error around
10 pb~\cite{CMS:2012iza}. We will use these experimental results to
constrain the $\phi$ parameter space in Section~\ref{sec:charmphysics}.

\subsection{$t\bar t$ pair-production}
We first comment on same-sign $t t$ production for this
model. Neglecting the coupling $\kappa_D$, $t$-number global symmetry is only broken by
the CKM elements and $t t$ production is suppressed. Including the
coupling $\kappa_D$, this model generates same-sign top final
state without CKM suppression, but with a large final state
multiplicity, for instance $uu\rightarrow tt\bar{b}\bar{b}dd$. The
cross section is then strongly suppressed by phase space. We therefore
concentrate on $t \bar t$ production. 

The dominant new physics contribution to $t \bar t$
pair-production is through exchange of the $\phi$ field in the $t$-channel.
In addition, the interference with SM gluon exchange in the $s$-channel cannot
be neglected.  The leading order parton-level pair production
cross-section, neglecting the mass of $\phi$, is given by
\beqa
\sigma(u\bar{u} \rightarrow t \bar{t}) &=& \frac{1}{216 \pi s} \left\{
\beta \left[
g_s^4 (8 m^2 + 4) - 6 g_s^2 ( 2 m^2 + 1) |\bar{\kappa}_{U_R}|^2 + 27 |\bar{\kappa}_{U_R}|^4
\right] \right. \nonumber \\
&& \left. \quad\quad\quad -  3 m^2 |\bar{\kappa}_{U_R}|^2 \ln{ \left( \frac{1 + \beta - 2 m^2}{1 - \beta - 2 m^2}  \right)  }
\left[ 4 g_s^2 ( m^2 + 1) - 9 |\bar{\kappa}_{U_R}|^2 \right]
\right\} \,,
\eeqa
where $m^2 \equiv m_t^2/s$ and $\beta = \sqrt{1- 4 m^2}$.
\begin{figure}[ht!]
\begin{center}
\includegraphics[width=0.6\textwidth]{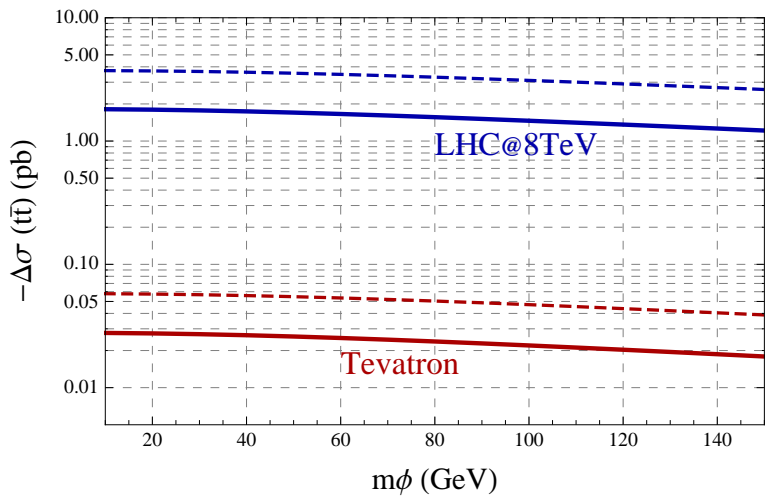} 
\caption{Modification of the $t\bar t$ production cross sections from contributions of the new neutral scalar $\phi$. The couplings are chosen to be $|\bar{\kappa}_{U_R}| = 0.2$ for the two solid lines and $|\bar{\kappa}_{U_R}| = 0.3$ for the two dashed lines. 
}
\label{fig:pairtopx}
\end{center}
\end{figure}
Figure~\ref{fig:pairtopx} shows the modifications on the $t\bar t$
production cross section.  Because of the deconstructive interference
(taking $|\bar{\kappa}_{U_R}|=0.2$ or 0.3), the new $\phi$
contribution decreases the total production cross section by a few pb
at the 8 TeV LHC and tens of fb at the Tevatron. The current uncertainty on
the $t\bar t$ pair-production cross section at the 8 TeV LHC is approximately
30~pb~\cite{ATLAS:2012gpa} from the ATLAS collaboration and
0.4~pb at Tevatron~\cite{Tevatronttbar}, so the modifications on $t\bar
t$ pair-production do not significantly constrain the parameter space
of this model.

\subsection{Non-Standard Top Decays}
\begin{figure}[ht!]
\begin{center}
\includegraphics[width=0.6\textwidth]{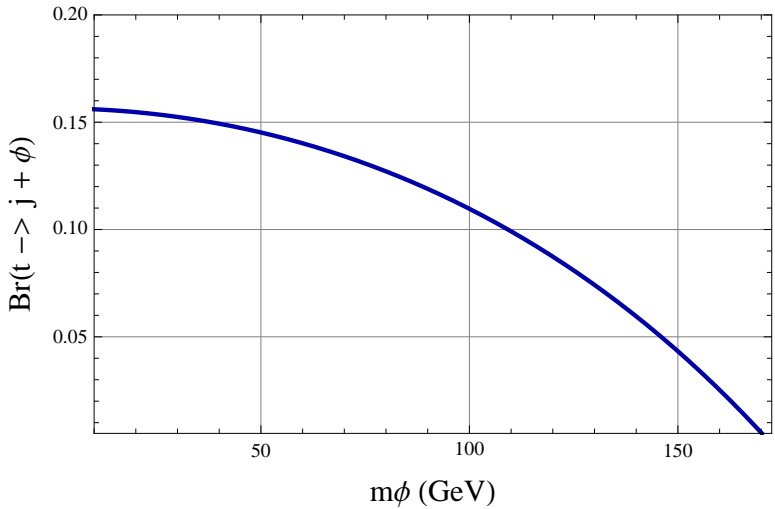} 
\caption{The branching ratio of the new top quark decay channel. The coupling is chosen to be $|\bar{\kappa}_{U_R}| = 0.2$. For a light $\phi$ field, the top quark may behave as a dijet at colliders. 
}
\label{fig:topdecay}
\end{center}
\end{figure}

For a light $\phi$ scalar with $m_\phi < m_t$, the top quark can decay
into $\phi$ plus the up quark or charm quark.  Since the CKM matrix is
nearly diagonal, top quark decay yields primarily $\phi_{13}$
($\phi_{23}$) for an up (charm) and $\phi$ final state.  A $\phi$ with
such flavor indices decays exclusively to a $b$ quark and a light
quark.  For $m_\phi \lesssim 
20$~GeV, the two resulting jets are collimated and
behave as a fat jet, so the top quark is observed as a dijet
resonance. For a heavier $\phi$, the top quark looks like a three-jet
resonance.  Summing the two dominant new
decay channels, $t \rightarrow u + \phi_{13}$ and $t \rightarrow c +
\phi_{23}$, the partial width for this mode is
\beqa
\Gamma(t \rightarrow j + \phi) = \frac{ |\bar{\kappa}_{U_R}|^2}{8\pi}\, m_t\,\left(1- \frac{m_\phi^2}{m_t^2}\right) \,.
\eeqa
Using the latest theoretical results for the top quark decay width in
the
SM~\cite{Czarnecki:1998qc,Chetyrkin:1999ju,Gao:2012ja,Brucherseifer:2013iv},
we show the branching ratio of the new decay channel in Fig.~\ref{fig:topdecay}. 
The current experimental value of the top quark width is $\Gamma_t =
1.99^{+0.69}_{-0.55}$~GeV from D0~\cite{Abazov:2010tm}, which is extracted
using the partial decay width $\Gamma(t \rightarrow Wb)$ measured
from the $t$-channel cross section for single top quark production
and from the branching fraction of ${\rm Br}(t \rightarrow W b)$
measured in $t\bar t$ events.  As we will see below, the total top quark width measurement
does constrain the parameter space for $\phi$.

The $\phi$ can also introduce additional contributions to the top
quark forward-backward or charge asymmetries. For the
parameter space allowed by the single top production cross section,
however, the modifications on the top quark pair production are negligible, and
hence the amount of asymmetry is unlikely to be visible at hadron
colliders. A lepton collider such as the International Linear Collider
\cite{AguilarSaavedra:2001rg,Weiglein:2004hn,Abe:2010aa,Aihara:2009ad}
would be an ideal machine to probe this model's parameter space.

\section{Additional constraints and D-meson physics}
\label{sec:charmphysics}
Both $b$-quark and $c$-quark physics may be modified by the introduction of the
new light scalar. As mentioned in the introduction, we are
interested in the up-type quark sector of MFV models.  The main
focus of our work is therefore on $\Delta T=1$ and $\Delta C=1$
processes.  On the other hand, the MFV mediator, $\phi$, also couples to
down-type quarks, modifying their properties.  We study the most
accessible $b$-quark and $c$-quark physics in this section.

\subsection{Enhanced $b$ Production}
\begin{figure}[!tb]
  \centering
  \begin{tikzpicture}
    \draw[gauge] (0,0) node[left] {$Z$} -- (2,0);
    \draw[fermion] (2,0) -- (3,1);
    \draw[fermion] (3,1) -- (4,2) node[right] {$q_i$};
    \draw[fermion] (4,-2) node[right] {$\overline{q_j}$} -- (2,0);
    \draw[scalar-ch] (4,0) node[right] {$\phi_{ji}$} -- (3,1);
  \end{tikzpicture}
  \hspace{1cm}
  \begin{tikzpicture}
    \draw[gauge] (0,0) node[left] {$Z$} -- (2,0);
    \draw[fermion] (2,0) -- (4,2) node[right] {$q_i$};
    \draw[fermion] (4,-2) node[right] {$\overline{q_j}$} -- (3,-1);
    \draw[fermion] (3,-1) -- (2,0);
    \draw[scalar-ch] (4,0) node[right] {$\phi_{ji}$} -- (3,-1);
  \end{tikzpicture}
  \caption{Diagrams for the decay $Z \to q_i \overline{q_j} \phi_{ji}$.}
  \label{fig:ZdecayFeyn}
\end{figure}
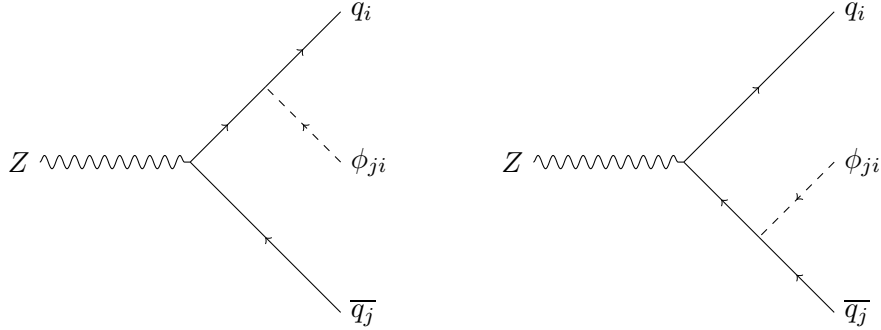

The interactions introduced in Eq.~(\ref{eq:phiInteraction})
yield unsuppressed couplings of $\phi$ to the down-type
sector. Fortunately, the coupling $\kappa_D$ does not
mediate flavor and does not allow for flavor violation without
coupling to the up sector, so many processes, including $\Delta F = 2$ 
FCNCs are not induced. There are subdominant bounds at low masses $m_\phi
\lesssim 10~{\rm GeV}$ from fixed-target searches for hadronic
resonances~\cite{Kephart:1977ad}. For the masses $m_\phi >10$~GeV
that we are studying,  the most stringent bounds come from $Z$
decays via a process illustrated in
Fig.~\ref{fig:ZdecayFeyn}. Existing LEP searches are most sensitive to
three body 
decays of the form $Z \to q\overline{q}^\prime \phi$.  In particular,
$Z \to b\overline{b} \phi$ is severely constrained from $Z \to b\bar{b}b\bar{b}$
searches~\cite{PDG}.  These searches require only three $b$-jets, under the
assumption that the fourth will be present in all cases.  A
$\phi_{33}$ would either look exactly like a $b$-jet at low mass or would decay
dominantly to two $b$-jets at higher masses.\footnote{The LEP
  experiments used a jet definition based on a cut on $y =
  M_{ij}^2/s$.  Typical values of $y_{\rm min}$ are few
  $\times~10^{-2}$, corresponding to $M_{ij} \sim 20~{\rm GeV}$.
  Using this definition, a relatively light $\phi$ would appear as a
  single jet to the experiment.}  Both final states would be accepted
by the analysis cuts for the $Z \to b\bar{b}b\bar{b}$ measurement.

%

To work out the bounds in detail, we also need to know the properties
of the $\phi$ field, in particular the branching ratio for its decay into two
$b$-jets.  Because the coupling $\kappa_D$ does not violate flavor,
the final state from $Z \rightarrow q\bar{q}^\prime \phi$ should
contain an even number of $b$-jets. Since only the decay with
$\phi_{33}$ can give 4$b$ final states, we only need to consider $Z
\rightarrow q\bar{q}^\prime \phi_{33}$, where the $\phi_{33}$ field mainly
decays into two $b$-jets. Other $\phi_{33}$ decay channels via its couplings to
up-type quarks are suppressed both by CKM angles and Yukawa couplings;
they can be neglected.  We therefore assume 100\% branching ratio for
$\phi_{33} \rightarrow b\overline{b}$.

To get a better idea of the constraints, we calculate the partial
width of the general processes $Z \to q_i \overline{q}_j \phi_{ji}$.  The rate $Z \to
\overline{q}_i q_j \phi^*_{ji}$ is the same by $CP$. The
details of our calculation of the partial width are given in Appendix
\ref{sec:appa}.  The integration over phase space was performed
numerically including all quark masses and the $\phi$ mass.
Nevertheless, it is instructive to examine an approximate expression for
the partial width in the limit $m_b, m_\phi \ll m_Z$
\begin{equation}
\Gamma(Z \to q_i \overline{q}_j \phi_{ji}) \sim \frac{\alpha \kappa_D^2
  m_Z}{576 \pi^2 s_w^2 c_w^2} = (0.35~{\rm MeV}) \times \kappa_D^2.
\end{equation}
This corresponds to a branching fraction of order $10^{-4}$ for
$\kappa_D$ of order $1$, which, as
we see below, is close to the current sensitivity.

As mentioned above, the most severe constraint arises from LEP searches
for $Z \to b\bar bb\bar b$.  The rate for this process is measured by
both the OPAL and DELPHI collaborations~\cite{Abreu:1999qh,
  Abbiendi:2000zt} with branching ratio
\begin{equation}
  {\rm Br}_{\rm exp}(Z \to b\bar{b}b\bar{b}) = (3.6 \pm 1.3) \times 10^{-4} \,.
\end{equation}
The total $Z$ width has been measured as $\Gamma_{\rm SM} =2.4952\pm0.0023$~GeV~\cite{PDG}.  The branching ratio
from the new physics with $m_\phi = 15$~GeV and $|\kappa_D|=1$  is 
\beqa
  {\rm Br}(Z \to b\overline{b}\phi_{33} \rightarrow 2b2\bar{b} ) + {\rm Br}(Z \to b\overline{b}\phi_{33}^* \rightarrow  2b2\bar{b}) = 2.9 \times 10^{-4}  \,.
\eeqa

\begin{figure}[th!b]
  \centering
  \includegraphics[width=0.6\textwidth]{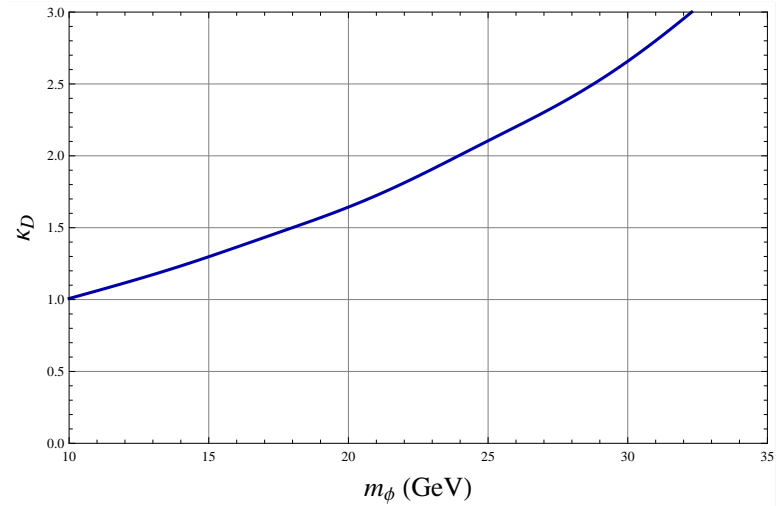}
  \caption{Upper limit on $\kappa_D$ from the measurement of $\rm{Br}(Z \to b\bar b \bar bb)$ as a function of $\phi$ particle masses.}
  \label{fig:bound-kD}
\end{figure}

Obtaining an accurate SM prediction for the $4b$ branching fraction of
the $Z$ is
challenging because it is a QCD process with large logs.  On the other
hand, it is certain that the new physics contribution cannot exceed
the upper limit on the total observed branching fraction. Using the
1$\sigma$ upper limit given above, $\rm{Br}_{\rm new}(Z \rightarrow
b\bar{b} b \bar{b}) < 4.9\times 10^{-4}$, the constraint on the
coupling for $m_\phi = 15~{\rm GeV}$ is  $\kappa_D < 1.28$. The bound
over the full range of interesting $\phi$ masses is presented in
Fig.~\ref{fig:bound-kD}. We can see from this figure that the limits on
the coupling $\kappa_D$ reaches the non-perturbative range for a
heavy $m_\phi\sim 30$~GeV. 

\subsection{$D$ Meson Direct $CP$ violation}
A striking consequence of a new light neutral scalar for charm physics
is the possibility of generating a significant direct $CP$ asymmetry
in $D$ meson decays.  In some parts of parameter space, the asymmetry
generated can be large enough to explain the anomaly observed by
Belle, CDF, and LHCb without requiring an enhancement of ``penguin''
contributions relative to the naive expectation, as explored in
Ref.~\cite{Isidori:2011qw}.  Other effects on
charm physics are negligible as they are overwhelmed by SM
contributions.  For example, $D^0-\overline{D^0}$ mixing typically
provides a very stringent bound for new physics contributions to the
$\Delta F=2$ processes.  With MFV implemented at tree-level, however,
$D^0-\overline{D^0}$ mixing is generated only at one loop with
bottom Yukawa suppression and does not constrain parameter
space of this model.

Direct $CP$ violation in the $D$ system has long been lauded as a
``smoking gun'' signature of BSM physics~\cite{Grossman:2006jg}.  Recent measurements of the
direct $CP$ asymmetry \emph{difference} between $D^0 \to K^+ K^-$ and
$D^0 \to \pi^+ \pi^-$ may provide the first hint of such $CP$ violation.  The observable is defined as
\beqa
  \Delta A_{CP} \equiv A^K_{CP} - A^\pi_{CP} \,.
\eeqa
For the MFV models considered so far, we have $A^K_{CP} \gg
A^\pi_{CP}$ and thus $ \Delta A_{CP} \approx A^K_{CP}$ since the asymmetry
for pions is Yukawa suppressed.  To tag the $D^0$, one can use the $\pi^+$
from $D^{*+}\rightarrow D^{0} + \pi^+$ or the muon from $B \rightarrow
D \mu X$.  For the $\pi^+$ tagging, and adding the errors in
quadrature, we obtain the value $\Delta A_{CP} = (-0.46\pm 0.13)$\%
averaged over the results from
BaBar~\cite{Aubert:2007if}, Belle~\cite{Ko:2012px},
CDF~\cite{Collaboration:2012qw} and
LHCb~\cite{LHCb-CONF-2013-003}. For the muon tagging, the latest
measurement from LHCb using 1.0 fb$^{-1}$ data at 7 TeV has $\Delta
A_{CP} = (0.49\pm 0.30\pm 0.14)$\%~\cite{Aaij:2013bra}, which has
opposite sign compared to the $\pi$-tagged result. Combining the
results from both tagging channels, the current world-averaged direct
$D$ meson $CP$ violation result is~\cite{HQAG}: 
\beqa
   \Delta A^{\rm exp}_{CP} = (-0.329 \pm 0.121 )\%\,,
   \label{eq:ACPexp}
\eeqa
which corresponds to a 2.7$\sigma$ significance. The SM prediction 
for this quantity is estimated to be smaller than ${\cal
  O}(10^{-3})$~\cite{Grossman:2006jg}. 

In addition to the recent decrease in the significance of the observed
$CP$ violation in $D$ decays, there has been renewed theoretical
study of direct $CP$ violation in the $D$ system.  An enhancement of
the relevant hadronic matrix element, analogous to the $\Delta I
=\frac{1}{2}$ rule in Kaon physics (see Ref.~\cite{Blum:2012uk,
  Boyle:2012ys} for recent Lattice calculations), may predict a larger
value of  $\Delta A_{CP}$, as pointed out in
Ref.~\cite{Golden:1989qx,  Atwood:2012ac}.   Recent work has shown that the assumption
of a large ``$\Delta U = 0$ rule'' for $D$ decays, i.e.\ that $\Delta
U = 0$ amplitudes receive a factor of $\sim 10$ enhancement compared to $\Delta
U =1$ amplitudes, can simultaneously explain several outstanding
puzzles in $D$ physics~\cite{Grossman:2012ry}.  Such an enhancement is
larger than naively expected from QCD estimations, but an
accurate calculation is beyond the reach of current techniques in the
$D$ system. 

\begin{figure}[!tb]
  \centering
  \begin{tikzpicture}
    \draw[fermion] (0,0) node [left] {$c$} -- (2,0);
    \draw[fermion] (2,0) -- (3.414,1.414) node [right] {$u$};
    \draw[scalar] (2,0) -- node [above right] {$\phi,\phi^*$}
    (3.414,-1.414);
    \draw[fermion] (4.828,0) node [right] {$\overline{q}$} -- (3.414,-1.414);
    \draw[fermion] (3.414,-1.414) -- (4.828,-2.828) node [right] {$q$};
  \end{tikzpicture}
  \caption{Diagram for the $D$ meson decay to $KK$ and $\pi\pi$ via
    the new scalar $\phi$.  The $KK$ and $\pi\pi$ modes correspond to
    $q = s$ and $q = d$ respectively.}
  \label{fig:dmesoncp}
\end{figure}
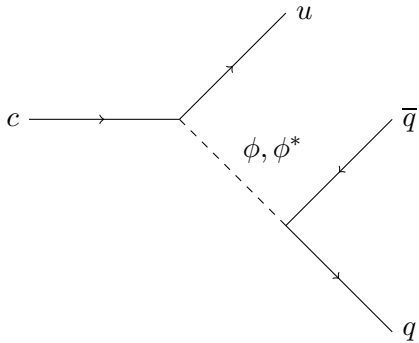

\begin{figure}[!tb]
  \centering
  \includegraphics[scale=0.8]{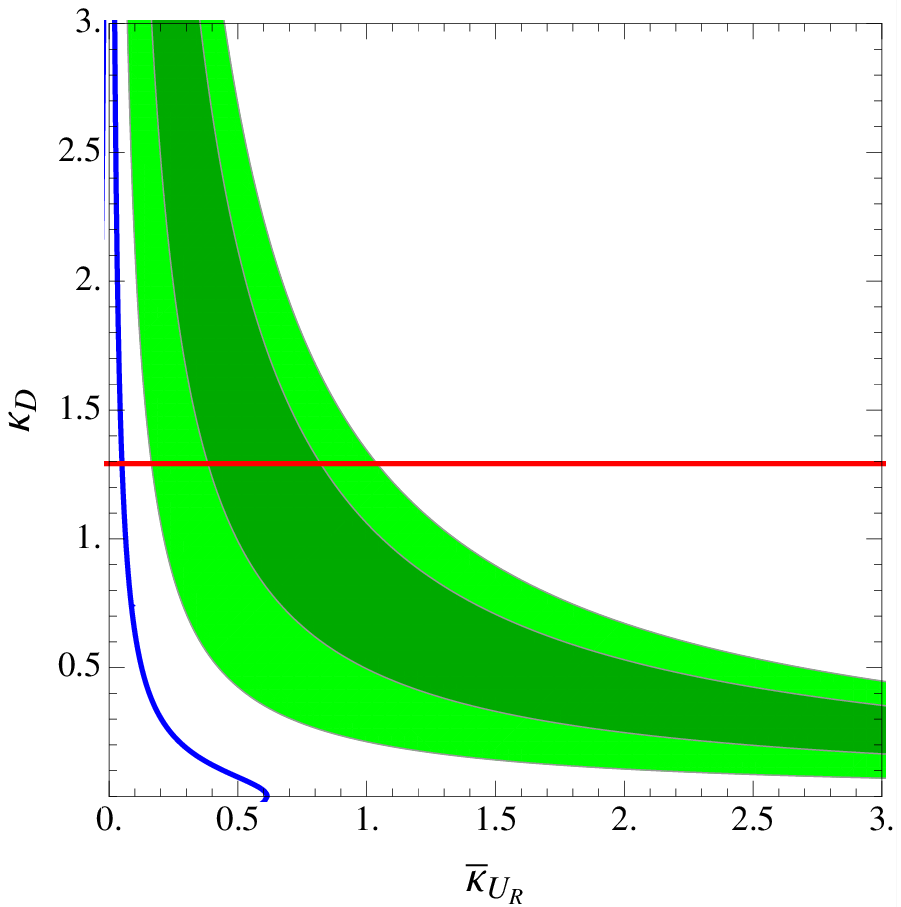} 
  \hspace{1.0cm}   
  \includegraphics[scale=0.8]{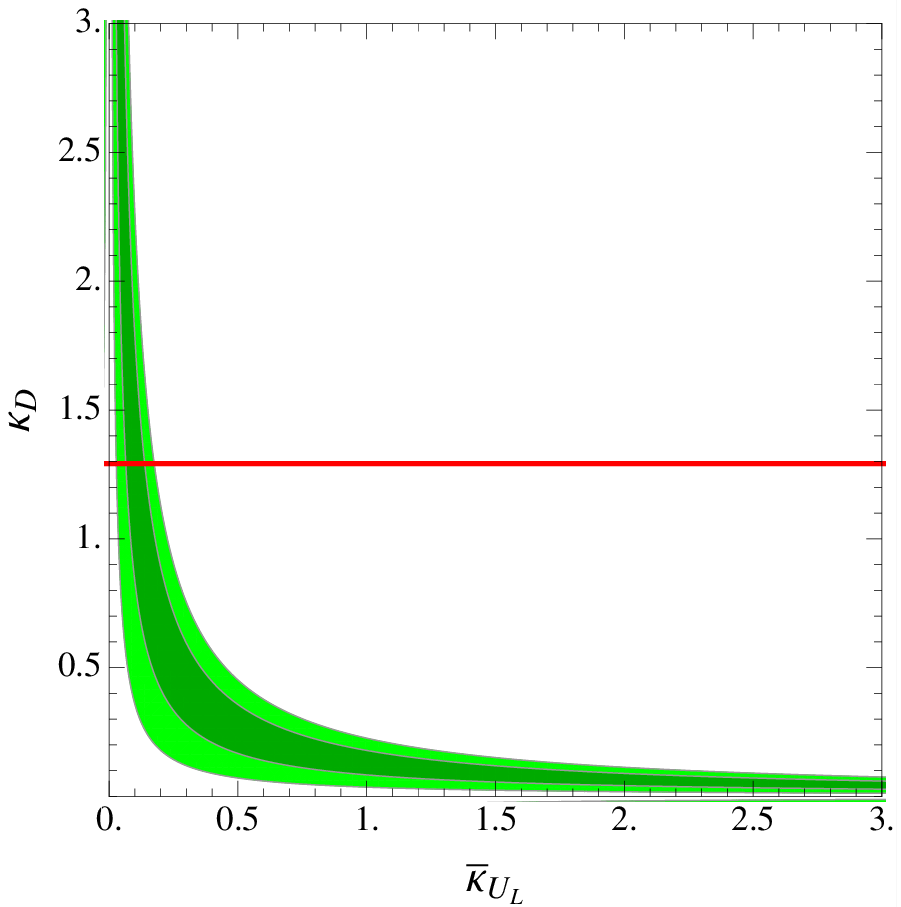}
  \caption{ The $1\sigma$ and $2\sigma$ contour plots of the parameter space to accommodate the direct $CP$ violation variable $A^{\rm exp}_{CP}$ in the $D$-meson system. The neutral scalar mass is chosen to be 10~GeV. The red horizontal line is the upper limit on $\kappa_D$ from the measurement of Br$(Z \rightarrow b\bar b b \bar b )$. The blue curve in the left panel is the upper limit constrained from the single-top production cross section at CDF.}
  \label{fig:dmesonconst}
\end{figure}

We now explore the possibility that the MFV operators of ${\cal
  O}_{V2}$ and ${\cal O}_{S2}$ explain the observed value of $\Delta A^{\rm
  exp}_{CP}$ in Eq.~(\ref{eq:ACPexp}).  We assume that naive 
factorization holds so that the SM contribution is negligible.  We
integrate out $\phi$ and run the resulting 
operators down to the $D$ meson mass scale.  The Feynman diagram 
for $D$ meson decay is illustrated in
Fig.~\ref{fig:dmesoncp}.  At the scale of the scalar $\phi$, we
generate the operators
\beqa
&&\frac{\kappa_{U_L} \kappa_D}{m_\phi^2} \, (\lambda_U^\dagger
V)_{il}\,(\lambda_D^\dagger V^\dagger)_{kj}\,  (\bar{u}^i_{R\,\alpha}
u^j_{L\,\alpha} ) (\bar{d}^k_{R\,\beta}  d^l_{L\,\beta}) \,+\, {\rm h.c.} \nonumber \\
&+& \frac{\kappa_{U_R} \kappa_D}{m_\phi^2} \, V_{il} \,
(\lambda_D^\dagger V^\dagger \lambda_U)_{kj}\,  (\bar{u}^i_{L\,\alpha}
u^j_{R\,\alpha} ) (\bar{d}^k_{R\,\beta}  d^l_{L\,\beta}) \,+\, {\rm h.c.}  \,. \label{eq:phiops}
\eeqa
%
%
From this equation, one can read off the Wilson coefficients:
\begin{equation}
  C_{V2}(m_\phi) = \frac{1}{2\,m_\phi^2} \kappa_{U_R} \kappa_D\,,\qquad C_{S2}(m_\phi) =
\frac{1}{2\,m_\phi^2} \kappa_{U_L} \kappa_D\,.
\end{equation}
In addition to these operators, a tensor operator is generated, but it
does not contribute to $CP$ violation as it has a zero matrix
element assuming naive factorization~\cite{Altmannshofer:2012ur}.  The
tensor operator does, however, give a significant contribution to the
renormalization group running of the scalar coefficients.  The details
of this running are described in Appendix~\ref{sec:appb}, while the
estimation of the relevant hadronic matrix element ratios are
performed in Appendix~\ref{sec:appc}.  In terms of the
low-energy operator, the direct $CP$ asymmetry is given by
\begin{equation}
  A_{CP}^{K} \approx \frac{2 \sqrt{2}}{N_c G_F} \lambda_c \lambda_s
  \left\{ \frac{1}{4} \sin{\delta_{V2}}\,  {\rm Im}\left[C_{V2}(m_D)\right] -
    \frac{1}{8}\, \chi_K \sin{\delta_{S2}} \,{\rm Im}\left[C_{S2}(m_D)\right] \right\} \,,
\end{equation}
where $\delta_{V2}$ and $\delta_{S2}$ are the strong phases of the matrix elements
of $\mathcal{O}_{V2}$ and $\mathcal{O}_{S2}$. One has a similar
expansion for $A_{CP}^{\pi}$ by replacing $\lambda_s$ by
$\lambda_d$. The strong phases are estimated to be $\mathcal{O}(1)$ 
in QCD decays and we take them to have the maximal value:
$\sin{\delta_{V2}}=1$ and $\sin{\delta_{S2}}=1$.  We also assume a
maximal weak phase for the coefficients of these operators: ${\rm
  arg}\,C_{V2} = \pi/2$ and ${\rm arg}\,C_{S2} = \pi/2$.  We neglect 
subdominant effects from interference between $\mathcal{O}_{V2}$ and 
$\mathcal{O}_{S2}$. The resulting regions of $\bar{\kappa}_{U_R}$--$\kappa_D$ and
$\bar{\kappa}_{U_L}$--$\kappa_D$ parameter space that accommodate the
$\Delta A_{CP}$ measurement are shown for $m_\phi
= 10~{\rm GeV}$ in Fig.~\ref{fig:dmesonconst}. In this figure, we
also show the constraints from Br$(Z \rightarrow
b\bar b \bar b b)$ and the single top quark production cross section
from CDF~\cite{CDFsingletop}, $\sigma^{\rm new}(\rm{single top}) <
3.61$~pb. We can see from the left panel of Fig.~\ref{fig:dmesonconst} that the entire
$A^{\rm exp}_{CP}$ preferred parameter space for
$\bar{\kappa}_{U_R}$--$\kappa_D$ has been excluded by the single top
production cross section measurement, while the right panel shows that
there is still allowed parameter space for $\bar{\kappa}_{U_L}$--$\kappa_D$. 

Before continuing, we turn to the question of whether there are any
further modes sensitive to the operators $\mathcal{O}_{V2}$
and/or $\mathcal{O}_{S2}$.  Several criteria must be satisfied for any
relevant observable.  Due to the Yukawa suppression of $D$ decays from
these operators, we consider only modes that are $CP$ violating to 
avoid competition with the dominant SM amplitude and involve $D \to K$
transitions to avoid additional Yukawa suppression.  Furthermore, we
have seen that $CP$ asymmetries generated in such transitions need to be
measured to fairly high precision.  There are
only a small number of observables that are close to satisfying all of
these criteria.  The most promising ones are $A_{CP}(D^0 \to
K^- \pi^+)$ ($\mathcal{O}_{V2}$), $A_{CP}(D^0 \to K^+
\pi^-)$ ($\mathcal{O}_{S2}$), and $A_{CP}(D^+ \to K^+ K^- \pi^+)$
($\mathcal{O}_{V2}, \mathcal{O}_{S2}$), where the operators in
parentheses yield significant sensitivity.  Modes
involving $K^0$ are challenging to compute as they receive dominant contributions
from $K$ -- $\bar{K}$ mixing, though they are among the most precisely
measured.  A full study of these additional observables is beyond the
scope of this work.

\section{Discussion and conclusions}
\label{sec:conclusion}
In this paper, we have studied a class of models below the scale of
electroweak symmetry breaking that lead to interesting $\Delta C = 1$
and $\Delta T = 1$ observables while remaining unconstrained by other
flavor and precision observables.  Despite these features, the models
cannot be the full story: they require a UV completion to render them
$SU(2)_L \times U(1)_Y$ invariant.  In order to have $\mathcal{O}(1)$
couplings in the IR model, new states must be introduced at a scale no
more than a factor of a few above the Higgs vacuum expectation value.  We now briefly outline
the general features of such UV completions.

There are only a limited number of possibilities in terms of
the gauge charges of the particle that completes the theory.  The
completion must involve a Higgs field in order to break electroweak
symmetry.  If the particle is a boson, then it must have the same
charges under SM gauge symmetries as a Higgs,\footnote{We assume that
  there are only $SU(2)_L$ doublet Higgses.  Other representations are
  highly constrained by data.} as it must couple the
relevant Higgs to the singlet $\phi$.  If the particle is a fermion,
then it must be a vector-like quark.  It may be either an $SU(2)_L$ doublet or singlet.
Thus, for the up- and down-sectors, there are three possibilities for the
field that implements the UV completion.  In the up-sector, they are:
  $\Phi(1,2)_{1/2}$, $\Psi_{L,R}(3,2)_{1/6}$, $\Psi_{L,R}(3,1)_{2/3}$,
  while in the down-sector, they are: $\Phi(1,2)_{1/2}$,
  $\Psi_{L,R}(3,2)_{1/6}$, $\Psi_{L,R}(3,1)_{-1/3}$. 
Note that in the first two cases the same particle may be responsible
for completing both the up- and the down-sectors. For each possible
set of gauge charges, there are several possible flavor charges.  The
phenomenology of the model depends greatly on the 
specific flavor charge, as well as any other degrees of freedom that
may appear near the scale of the UV completion.  A complete study is
beyond the scope of this work.

In summary, we have explored the phenomenology of the up-type quark
sector in the MFV framework. Concentrating on the $CP$ violating
effective operators, we have found interesting correlations
between the $\Delta C=1$ and $\Delta T =1$ processes. If the flavor
mediator has a mass below ${\cal O}(100~\rm{GeV})$, both processes
can be probed at the LHC either in flavor physics related to $D$-meson
decays or in the top quark physics related to top quark single and
pair production. A further consequence of the
existence of the light flavor mediator is a new decay channel for the
top quark. This new decay channel, $t\rightarrow c + \phi$, would not
appear in existing flavor-changing-neutral-current decay searches such
as $t\rightarrow c + Z$ because the top quark would appear as a
``dijet" resonance for a $\phi$ mass below ${\cal O}(20~\rm{GeV})$.
A fat jet analysis for the $\phi$ field from the top decay can
therefore probe the up-quark sector.

\subsection*{Acknowledgments} 
We would like to thank Amarjit Soni and Jure Zupan for useful discussions and comments. 
Y. Bai is supported by startup funds from the UW-Madison. SLAC is operated by Stanford University for the US Department of Energy under contract DE-AC02-76SF00515.

\begin{appendix}
\section{Calculation of the Partial Width $\Gamma(Z \to q \overline{q}^\prime \phi)$}
\label{sec:appa}

The tree-level diagrams for this decay are shown in
Fig.~\ref{fig:ZdecayFeyn}.  Note that there is a soft divergence in
the limit $m_\phi = m_q = m_{q^\prime} = 0$, so the calculation cannot
be done in the limit where the final state masses all vanish.  In this
Appendix, we present the differential width contributions for the
decay, as well as the integrated width assuming that the quark masses
vanish.  Note that the assumption of $m_q = 0$ is not necessarily
sufficiently accurate for final states involving the $b$ quark due to
the fact that $m_\phi$ is not much greater than $m_b$. 

The amplitudes for the decay are given by
\begin{equation}
  \label{eq:4}
  \mathcal{M}_1 = \overline{u}_i (i \kappa_{ij} P_R) \frac{i(\slashed{p}_1 + \slashed{p}_2) + m_j}{(p_1 + p_2)^2 - m_j^2} \slashed{\epsilon} (i g_L P_L + i g_R P_R) v_j,
\end{equation}
\begin{equation}
  \label{eq:5}
  \mathcal{M}_2 = \overline{u}_i \slashed{\epsilon} (i g_L P_L + i g_R P_R) \frac{i(\slashed{p}_2 + \slashed{p}_3) + m_i}{(p_2 + p_3)^2 - m_i^2} (i \kappa_{ij} P_R) v_j,
\end{equation}
where $\kappa_{ij}$ is the coupling of quarks $i$ and $j$ to the scalar $\phi_{ji}$, including all factors of Yukawas, $g_{L,R}$ are the couplings of left-handed and right-handed quarks $q_j$ to $Z$, and $p_{1,2,3}$ are the momenta of $q_i$, $\phi_{ji}$, and $\overline{q_j}$ respectively.

The resulting differential width is given by
\begin{equation}
  \label{eq:6}
  d\Gamma = \frac{1}{(2 \pi)^3} \frac{1}{32 m_Z^3} \overline{|\mathcal{M}|^2} d m_{12}^2 d m_{23}^2,
\end{equation}
\begin{align}
  \label{eq:6}
  \overline{|\mathcal{M}|^2} & = \left(-g_{\mu\nu} + \frac{p_\mu p_\nu}{m_Z^2}\right) |\kappa_{ij}|^2 \nonumber\\
  ~ & \left\{\frac{{\rm Tr}[(\slashed{p}_1 + m_i) P_R (\slashed{q}_1 + m_j) \gamma^\mu (g_L P_L + g_R P_R) (\slashed{p}_3 + m_j) (g_L P_R + g_R P_L) \gamma^\nu (\slashed{q}_1  + m_j) P_L]}{(m_{12}^2 - m_j^2)^2}\right. \nonumber\\
 ~ & \frac{{\rm Tr}[(\slashed{p}_1 + m_i) P_R (\slashed{q}_1 + m_j) \gamma^\mu (g_L P_L + g_R P_R) (\slashed{p}_3 + m_j) P_L \gamma^\nu (\slashed{q}_2+ m_i)  \gamma^\nu (g_L P_L + g_R P_R) ]}{(m_{12}^2 - m_j^2) (m_{23}^2 - m_i^2)}\nonumber\\
~ & \frac{{\rm Tr}[(\slashed{p}_1 + m_i) P_R (\slashed{q}_2 + m_i) \gamma^\mu (g_L P_L + g_R P_R) (\slashed{p}_3 + m_j) (g_L P_R + g_R P_L) \gamma^\nu (\slashed{q}_1  + m_j) P_L]}{(m_{12}^2 - m_j^2) (m_{23}^2 - m_i^2)}  \nonumber\\
~ & \left.\frac{{\rm Tr}[(\slashed{p}_1 + m_i)  P_R (\slashed{q}_2 + m_i) \gamma^\mu (g_L P_L + g_R P_R) (\slashed{p}_3 + m_j)P_L \gamma^\nu (\slashed{q}_2 + m_i)  \gamma^\nu (g_L P_L + g_R P_R)]}{(m_{23}^2 -  m_i^2)^2} \right\},
\end{align}
where $q_1 = p_1 + p_2$ and $q_2 = p_2 + p_3$.

The phase space integration yields an even more involved expression,
but in the limit $m_i = m_j = 0$, a relatively compact result emerges:
\begin{align}
\Gamma(Z \to q_i \overline{q}_j \phi_{ji}) & = \frac{\alpha
  \kappa_{ij}^2 m_Z}{576 \pi^2 c_w^2 s_w^2} \bigg\{(g_L^2 + g_R^2)
  \left[-17 + 9 x  + 9 x^2 - x^3 - 6 \log x - 18 x \log x\right]\nonumber\\~ &  + g_L
  g_R \bigg[10 +78 x - 90 x^2 + 2 x^3 + 60 x \log x + 36 x^2 \log x + 12
  x^2 \log^2 x\nonumber\\ ~& -48 x^2 \log x \log (1+x) -24 x^2
  \left(2 {\rm Li}_2(-x) +\frac{\pi^2}{6}\right)\bigg] \bigg\}\, ,
\end{align}
where $g_{L,R} = T_3 - Q s_w^2$ and $x = m_\phi^2/m_Z^2$.  This result
provides a good approximation of the result in the massive quark case
for physical quark masses and for $m_\phi \gtrsim 10~{\rm GeV}$.

\section{Wilson Coefficient Running}
\label{sec:appb}
In this Appendix, we present the details of the running of the Wilson
coefficients for the various effective operators.  In particular, we
determine the one-loop anomalous dimension matrices for the
coefficients of operators $\mathcal{O}_{V2}$ and $\mathcal{O}_{S2}$.
The anomalous dimension matrices are defined such that 
\begin{equation}
  \label{eq:2}
  \frac{dC_i}{d\mu} = \frac{\alpha_s}{4\pi} \gamma^T C_i \,,
\end{equation}
where $C_i$'s are the Wilson coefficients of the set of operators that mix.

The operator $\mathcal{O}_{V2}$ runs much like the operator
$\mathcal{O}^{(1)p}_2$ in \cite{Altmannshofer:2012ur} except that it
cannot receive contributions from penguin-like operators: those
operators have a distinct chiral structure and mixing is forbidden by
Lorentz symmetry.  The only mixing is therefore with
$\mathcal{O}_{V1}$.  The anomalous dimension matrix can be read off of
the upper left $2 \times 2$ block of the anomalous dimension matrix
for the operators $O_i^{(1)}$ in \cite{Altmannshofer:2012ur}: 
\begin{equation}
  \label{eq:1}
  \gamma_V = \begin{pmatrix} -\frac{6}{N_c} & 6 \\ 6 & -\frac{6}{N_c} \end{pmatrix}.
\end{equation}

The operator $\mathcal{O}_{S2}$ is identical in Lorentz and color
structure to the operator $\mathcal{O}^{(1)}_{S2}$ of
\cite{Altmannshofer:2012ur} and therefore the anomalous dimension
matrix is
\begin{equation}
  \label{eq:9}
  \gamma_S = \begin{pmatrix} \frac{6 - 6 N_c^2}{N_c}  & 0 & \frac{1}{N_c} & -1 \\
    -6 & \frac{6}{N_c} & -\frac{1}{2} & \frac{2 - N_c^2}{2 N_c} \\
    \frac{48}{N_c} & -48 & \frac{2 N_c^2 - 2}{N_c} & 0 \\
    -24 & \frac{48 - 24 N_c^2}{N_c} & 6 & \frac{4 N_c^2 + 2}{- N_c}\end{pmatrix}  \,.
\end{equation}

\section{Hadronic Matrix Element Estimation}
\label{sec:appc}

Recent work~\cite{Brod:2012ud,Grossman:2012ry} has demonstrated a
consistent picture for observed $D$ meson physics within the SM under
the assumption of a large deviation from naive factorization.  On the
other hand, this picture has yet to be confirmed by direct calculation
of the hadronic matrix elements.  It remains possible that there is
enhanced $CP$ violation in the $D$ system due solely or partly to new
physics contributions.  The results regarding $D$ meson $CP$ violation
in this paper therefore assume that naive factorization gives a
reasonable estimate of the relative sizes of the various hadronic
matrix elements contributing to $D$ meson decays.  In this Appendix,
we present the details of the estimation used to calculate $\Delta
A_{CP}$ in this paper, following the work of
Ref.~\cite{Altmannshofer:2012ur}.  

Naive factorization is the assumption that a hadronic matrix element
$\langle h^+ h^-| (\overline{u} \Gamma_1 q) (\overline{q} \Gamma_2 c)
|D^0\rangle$ can be reliably estimated by 
\begin{equation}
  \label{eq:fact}
  \langle h^+ h^-| (\overline{u} \Gamma_1 q) (\overline{q} \Gamma_2 c) |D^0\rangle \approx \langle h^+| (\overline{u} \Gamma_1 p)|0\rangle\langle h^-| (\overline{p} \Gamma_2 c) |D^0\rangle.
\end{equation}
Using \eqref{eq:fact}, we can relate the hadronic matrix elements for
the operators $\mathcal{O}_{V2}$ and $\mathcal{O}_{S2}$ to that for
the leading SM operator ${\cal O}_{\rm SM}\equiv (\overline{u}_L
\gamma^\mu q_L) (\overline{q}_L \gamma_\mu c_L)$.  Under
factorization, we can write: 
\begin{equation}
  \label{eq:7}
  \langle h^+ h^- | (\overline{u}_{L\alpha} \gamma^\mu q_{L\alpha}) (\overline{q}_{L\beta} \gamma_\mu c_{L\beta}) |D^0\rangle \approx -\frac{i}{2} \delta_{\alpha\alpha} \delta_{\beta\beta}\, p_{h^+} \cdot p_{h^-} f_h\, f_{+}^{Dh^-} = -\frac{i}{2} N_c^2\, p_{h^+} \cdot p_{h^-} f_h\, f_{+}^{Dh^-},
\end{equation}
where we define $M_1^\mu = \langle h^+| (\overline{u}_{L\alpha}
\gamma^\mu q_{L\alpha})|0\rangle$ (no sum over $\alpha$) and $M_2^\mu
= \langle h^-| (\overline{q}_{L\alpha} \gamma^\mu c_{L\alpha})
|D^0\rangle$ (no sum over $\alpha$). 
Similarly, for the quark part of $\mathcal{O}_{V2}$, we find
\begin{equation}
  \label{eq:8}
  \langle h^+ h^- | (\overline{u}_{L\alpha} \gamma^\mu q_{L\beta}) (\overline{q}_{R\beta} \gamma_\mu c_{R\alpha}) |D^0\rangle \approx - \frac{i}{2} \delta_{\alpha\beta} \delta_{\alpha\beta}\, p_{h^+} \cdot p_{h^-} \,f_h\, f_{+}^{Dh^-}= -\frac{i N_c}{2} p_{h^+} \cdot p_{h^-} \,f_h\, f_{+}^{Dh^-}.
\end{equation}
Finally, we consider $\mathcal{O}_{S2}$:
\begin{align}
  \label{eq:8}
  \langle h^+ h^- | (\overline{u}_{L\alpha} q_{R\beta}) (\overline{p}_{L\beta} c_{R\alpha}) |D^0\rangle & \approx \frac{i}{2} \delta_{\alpha\beta} \delta_{\alpha\beta} \frac{m_h^2}{m_q + m_u} f_h \frac{(p_D - p_{h^-}) \cdot p_{h^-}}{m_c - m_s} f_{+}^{Dh^-} \nonumber\\
  ~ & \approx i\, \frac{N_c}{2}\frac{m_h^2}{m_c (m_q + m_u)} \,p_{h^+} \cdot p_{h^-}\, f_h\, f_{+}^{Dh^-}\,.
\end{align}
Note that we relate the (pseudo-)scalar matrix elements to the (pseudo-)vector matrix elements using the Dirac equation.  From these results, we obtain the following relations, assuming naive factorization:
\begin{equation}
  \label{eq:3}
  \mathcal{O}_{\rm SM} = N_c\, \mathcal{O}_{V2} = \frac{2 N_c}{\chi_f}  \mathcal{O}_{S2},
\end{equation}
where $\chi_{K} \approx 2 m_K^2/[m_c (m_s + m_u)]\approx 4.2$ and $\chi_{\pi} \approx 2 m_\pi^2/[m_c (m_d + m_u)]\approx 2.8$.

\end{appendix}

\bibliography{UpSectorMFV}
\bibliographystyle{JHEP}
 \end{document}